\documentclass[twocolumn,preprintnumbers,amsmath,amssymb]{revtex4}

\usepackage{graphicx}% Include figure files
\usepackage{dcolumn}% Align table columns on decimal point
\usepackage{bm}% bold math
\usepackage[usenames]{color}    % comments and corrections in colors

\usepackage{graphicx}% Include figure files
\usepackage{bm}% bold math
\usepackage{color}
\def\>{\rangle}\def\<{\langle}
\def\togli#1{}
\def\comment#1{}
\def\labell#1{\label{#1}}
\begin{document}
%\fbox{{\scriptsize Preliminary draft. \today.}}  

\title{Time from quantum entanglement: 
an experimental illustration}
\author{Ekaterina Moreva$^1,2$, Giorgio Brida$^1$, Marco Gramegna$^1$, Vittorio Giovannetti$^3$, Lorenzo Maccone$^4$,  Marco Genovese$^1$}
\affiliation{\vbox{$^1$INRIM, strada delle Cacce 91, 10135 Torino,
    Italy} \vbox{$^2$International Laser Center of M.V.Lomonosov Moscow State University, 119991, Moscow, Russia}\vbox{$^3$NEST, Scuola Normale Superiore and Istituto
    Nanoscienze-CNR, piazza dei Cavalieri 7, I-56126 Pisa, Italy}
\vbox{$^3$Dip.~Fisica ``A.~Volta'', INFN Sez.~Pavia, Univ.~of Pavia,
  via Bassi 6, I-27100 Pavia, Italy}}
\begin{abstract} In the last years several theoretical papers discussed if time can be an emergent propertiy deriving from quantum correlations.
Here, to provide an insight into how this phenomenon can occur,  we present an experiment that  illustrates Page and
  Wootters' mechanism of ``static'' time, and Gambini et al.
  subsequent refinements. A static, entangled state between a clock
  system and the rest of the universe is perceived as evolving by
  internal observers that test the correlations between the two
  subsystems. We implement this mechanism using an entangled state of
  the polarization of two photons, one of which is used as a clock to
  gauge the evolution of the second: an ``internal'' observer that
  becomes correlated with the clock photon sees the other system
  evolve, while an ``external'' observer that only observes global
  properties of the two photons can prove it is static.
\end{abstract}
\pacs{}
\maketitle

{\flushright{\em ``Quid est ergo tempus? si nemo ex me quaerat, scio;
    si quaerenti explicare velim, nescio.'' \cite{a}}}
\vskip.45\baselineskip
\togli{{\flushright{\em ``...an infinite series of times, in a dizzily growing, ever
  spreading network of diverging, converging and parallel times. This
  web of time--the strands of which approach one another, bifurcate,
  intersect or ignore each other through the centuries--embraces
  every possibility.'' \cite{borges}}}}

{ The ``problem of time''
  \cite{kuchar,isham,ashtekar,anderson,altra} in essence stems from
  the fact that a canonical quantization of general relativity yields
  the Wheeler-De Witt equation \cite{wdw,hartle} predicting a static
  state of the universe, contrary to obvious everyday evidence. A
  solution was proposed by Page and Wootters \cite{pw,w}: thanks to
  quantum entanglement, a static system may describe an evolving 
  ``universe'' from the point of view of the internal observers.
  Energy-entanglement between a ``clock'' system and the rest of the
  universe can yield a stationary state for an (hypothetical) external
  observer that is able to test the entanglement vs.  abstract
  coordinate time. The same state will be, instead, evolving for
  internal observers that test the correlations between the clock and
  the rest \cite{pw,w,pagereply,gambinipullin,peresamjphys48,rovelli}.
  Thus, time would be an emergent property of subsystems of the
  universe deriving from their entangled nature: an extremely elegant
  but controversial idea \cite{kuchar,unruh}. Here we want to
  demystify it by showing experimentally that it can be naturally
  embedded into (small) subsystems of the universe, where Page and
  Wootters' mechanism (and Gambini et al. subsequent refinements
  \cite{gambinipullin,montev}) can be easily studied. We show how a
  static, entangled state of two photons can be seen as evolving by an
  observer that uses one of the two photons as a clock to gauge the
  time-evolution of the other photon. However, an external observer
  can show that the global entangled state does not evolve.}

Even though it revolutionizes our ideas on time, Page and Wootters'
(PaW) mechanism is quite simple \cite{pw,w,pagereply}: they provide a
static entangled state $|\Psi\rangle$ whose subsystems evolve according to the
Schr\"odinger equation for an observer that uses one of the subsystems
as a clock system $C$ to gauge the time evolution of the rest $R$.   
While the division into subsystems is largely arbitrary,
the PaW model 
 %and, as long as  the interactions among them  can be neglected
  %the global Hamiltonian of the model
   assumes the possibility of neglecting interaction among them writing the Hamiltonian of the global system as  ${\cal H}={\cal H}_c\otimes\openone_r+\openone_c\otimes
{\cal H}_r$, where ${\cal H}_c,{\cal H}_r$ are the local terms associated with $C$ and $R$, respectively \cite{w}.
In this framework
the state of the ``universe''  $|\Psi\rangle$ is then identified
by  enforcing the  
Wheeler-De Witt equation
${\cal H}|\Psi\>=0$, i.e. 
by requiring  $|\Psi\rangle$ to be  an eigenstate of ${\cal H}$ for the zero eigenvalue. The rational of this choice follows from
the observation that by projecting $|\Psi\rangle$
on the states  $|\phi(t)\rangle_c= e^{-i H_c t/\hbar} |\phi(0)\rangle_c$
of the clock, one gets  the vectors 
%%%%%%%%%%%%%%%%%%%%%%%%

\begin{figure}[tbp] \begin{center}
\includegraphics[width=0.5\textwidth]{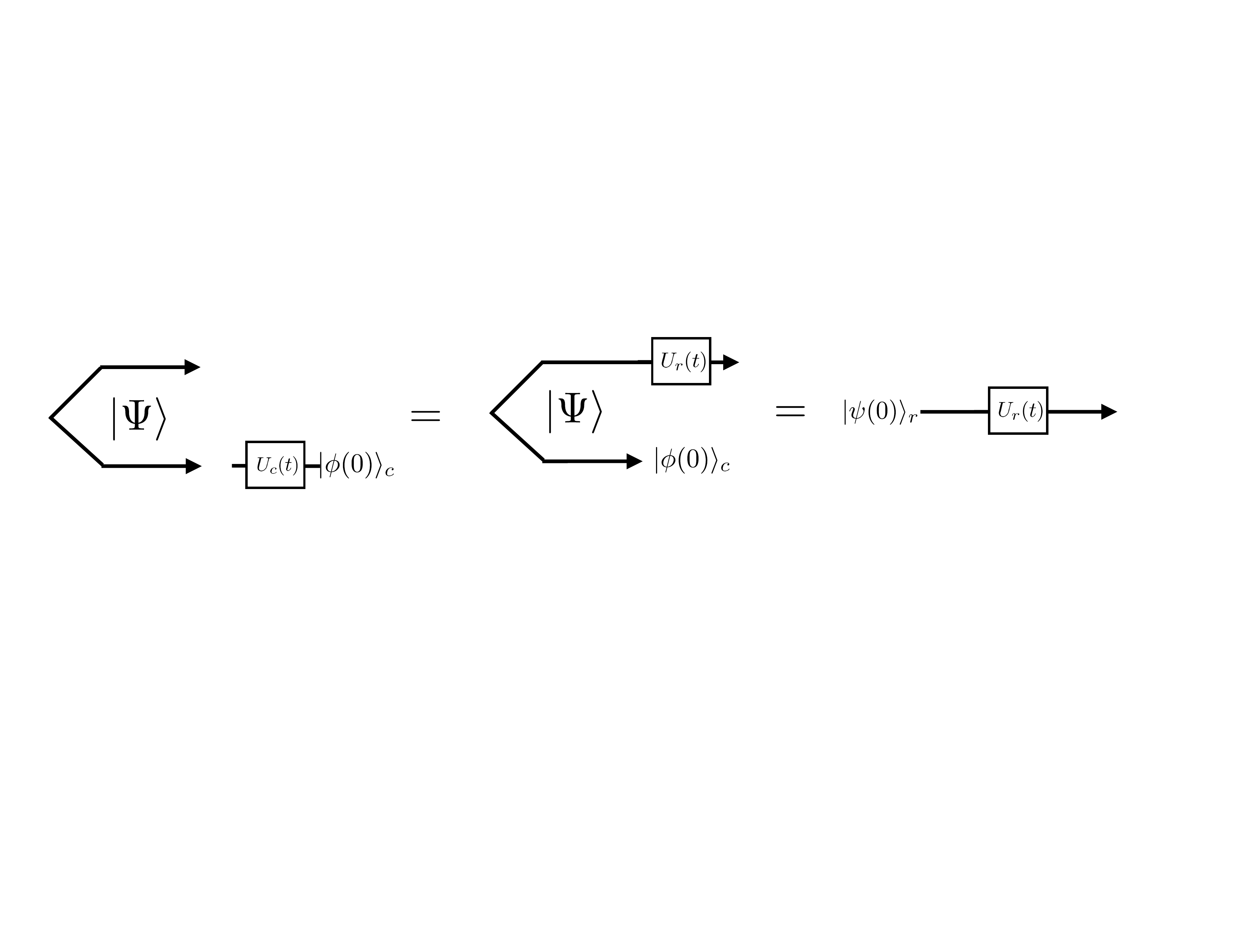}

\caption{Gate array  representation of the PaW mechanisms~\cite{pw,w,pagereply} for a CR non interacting model. Here
$U_r(t) = e^{-i H_rt/\hbar}$ and $U_c(t) = e^{-i H_ct/\hbar}$
are the unitary time evolution operators of the clock
$C$ and of the rest of universe $R$ respectively.
$|\Psi\rangle$ is the global state of the system which 
is assumed to be eigenstate with null eigenvalue of the
global Hamiltonian $H=H_c + H_r$ (see text).}
\label{f:figu0} \end{center} \end{figure}

%%%%%%%%%%%%%%%%%%%%%%%%%%%
\begin{eqnarray} \label{prima}
|\psi(t)\rangle_r :=  {_c\langle} \phi(t)|\Psi\rangle =
e^{- i H_r t/\hbar} |\psi(0)\rangle_r \;,
\end{eqnarray} 
 that describe a proper evolution 
of the subsystem $R$ under the action of its local Hamiltonian $H_r$, the initial state being $|\psi(0)\rangle_r = {_c\langle} \phi(0)|\Psi\rangle$ (see
Fig.~\ref{f:figu0}).
Therefore, despite the fact that globally the system
appears to be static, its components exhibits 
correlations that mimics the presence of a dynamical evolution \cite{pw,w,pagereply}. 
 Two main flaws of the PaW mechanisms  have
been pointed out \cite{unruh,kuchar}. The first is based on  the (reasonable) skepticism
 to accept that quantum mechanics may describe a system as
large as the universe, together with its internal observers
\cite{pagereply,gambinipullin}.  The second has a more practical character and is based on the
observation that in the PaW model the calculations of transition probabilities and of
propagators appears to be  problematic~\cite{kuchar,pagereply}.
An attempt to fix the latter issue has been discussed  by Gambini et al.~(GPPT) \cite{gambinipullin,montev}
by extending  a proposal by Page
\cite{pagereply} and invoking the notion of
`evolving constants' of Rovelli~\cite{rovt} (a brief
overview of this approach is given in the appendix).

\begin{figure}[tbp] \begin{center}
\includegraphics[width=0.5\textwidth]{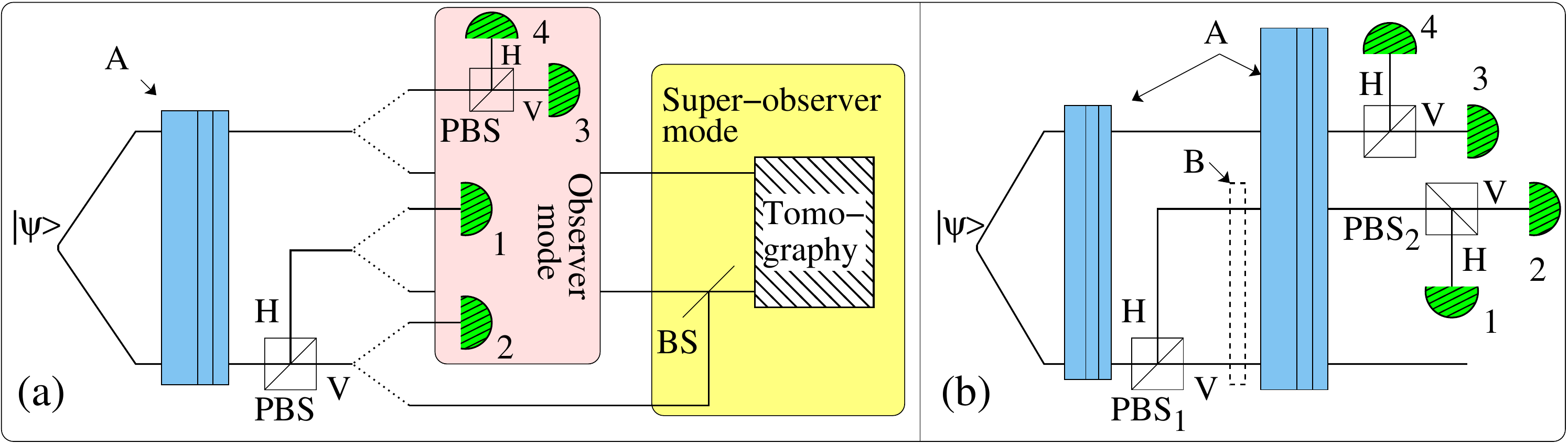}
\caption{Details of the experiment. {\bf (a)}~``Observer'' and
  ``super-observer'' mode in the PaW mechanism: one subsystem
  (polarization of the upper photon) evolves with respect to a clock
  constituted by the other subsystem (polarization of the lower
  photon). The experimenter in observer mode (pink box) can prove the
  time evolution of the first photon using only correlation
  measurements between it and the clock photon without access to an
  external clock. The super-observer mode (yellow box) proves through
  state tomography that the global state of the system is static. {\bf
    (b)}~Two-time measurements in the GPPT mechanism: the two time
  measurements are represented by the two polarizing beam splitters
  PBS$_1$ and PBS$_2$ respectively.  The blue boxes (A) represent
  different thicknesses of birefringent plates which evolve the
  photons by rotating their polarization: different thicknesses
  represent different time evolutions. The PaW mechanism (a) is
  completely independent of the thickness, whereas the GPPT mechanism
  (b) allows it to be measured by the experimenter only through the
  clock photon (the abstract coordinate time is unaccessible and
  averaged away); the dashed box (B) represents a (known) phase delay
  of the clock photon only; PBS stands for polarizing beam splitter in
  the $H/V$ basis; BS for beam splitter.}
\label{f:schema} \end{center} \end{figure}

In this work we present an experiment which 
allows reproducing the basic features of the PaW and GPPT models. In particular the PaW model is
realized by identifying $|\Psi\>$ with  
an  entangled state of the vertical $V$ and horizontal $H$ polarization degree of freedom of two photons in two spatial modes
$c,r$, i.e.  (see following section)
\begin{eqnarray}
  |\Psi\>=\tfrac1{\sqrt{2}}(|H\>_c|V\>_r-|V\>_c|H\>_r)
 \labell{st}\;,
\end{eqnarray}
and enforcing the Wheeler-De Witt equation by taking 
 ${\cal H}_c={\cal H}_r=i\hbar\omega(|H\>\<V|-|V\>\<H|)$ 
 as local Hamiltonians of the system ($\omega$ being a
 parameter which defines the time scale of the model).
 For this purpose rotations of the
   polarization 
 of the two
  photons are induced by forcing them to
  travel through identical  birefringent plates as shown in Fig.~\ref{f:schema}.
This allows us to consider a setting where everything can
  be decoupled from 
the ``flow of time'', i.e.~when the photons are traveling outside the
plates.  Nonetheless, the clock photon is a true (albeit extremely
simple) clock: its polarization rotation is proportional to the time
it spends crossing the plates.
\par
Although extremely
simple, our model
captures the two, seemingly contradictory, properties of the PaW
mechanism:  the evolution of the subsystems relative to each other, and
the staticity of the global system. This is achieved by running the
experiment in two different modes (see Fig.~\ref{f:schema}a): (1)~an
``observer'' mode, where the experimenter uses the readings of the
clock photon to gauge the evolution of the other: by measuring the
clock photon polarization he becomes correlated with the subsystems
and can determine their evolution. This mode describes the
conventional observers in the PaW mechanism: they are, themselves,
subsystems of the universe and become entangled with the clock systems
so that they see an evolving universe; (2)~a ``super-observer'' mode,
where he carefully avoids measuring the properties of the subsystems
of the entangled state, but only global properties: he can then
determine that the global system is static. This mode describes what
an (hypothetical) observer external to the universe would see by
measuring global properties of the state $|\Psi\>$: such an observer
has access to abstract coordinate time (namely, in our experimental
implementation he can measure the thickness of the plates) and he
can prove that the global state is static, as it will not evolve even
when the thickness of the plates is varied.

\begin{figure}[tbp] \begin{center}
\includegraphics[width=0.5\textwidth]{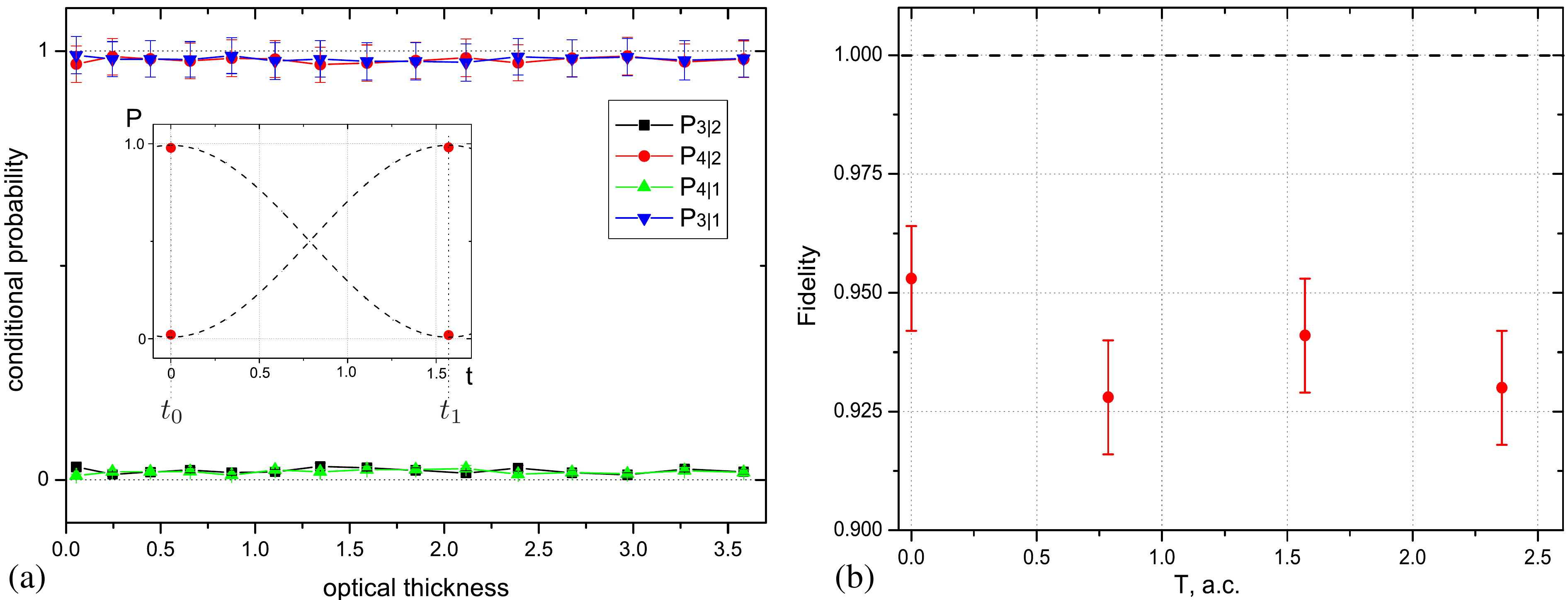}
\caption{PaW experimental results.  (a)~Observer mode: plot of
    the clock-time dependent probabilities of measurement outcomes as
    a function of the of the plate  thickness (corresponding to abstract coordinate time $T$): circles and
    squares represent $p(t_1)=P_{3|1}$ and $p(t_2)=P_{3|2}$
    respectively, namely the probabilities of measuring $V$ on the
   subsystem 1 as a function of the clock time $t_1$, $t_2$; circles and
    triangles represent $P_{4|1}$ and $P_{4|2}$, the probabilities of
    measuring $H$ on the subsystem 1 as a function of the clock time. As
    expected from the PaW mechanism, these probabilities are
    independent of the abstract coordinate time $T$, represented by
    different phase plate A thicknesses (here we used a 957$\mu$m
    thick quartz plate rotated by 15 different equiseparated angles).  The inset shows the graph that the
    observer himself would plot as a function of clock-time: circles
    representing the probabilities of finding the system photon $V$ at
    the two times $t_1$, $t_2$, the triangles of finding it $H$.
    (b)~Super-observer mode: plot of the conditional fidelity between
    the tomographic reconstructed state and the theoretical initial
    state $|\Psi\>$ of Eq.~\eqref{st} as a function of the abstract coordinate time $T$.  The
    fidelity $F = \<\Psi|\rho_{out}|\Psi\>$ (which measures the
    overlap between the theoretical initial state $|\Psi\>$ and the
    final state $\rho_{out}$ after its evolution through the plates)
    is conditioned on the clock photon exiting the right port of the
    beam splitter BS. The fact that the fidelity is constant and close
    to one (up to experimental imperfections) proves that the global
    entangled state is static. }
\label{f:results} \end{center} \end{figure}

In observer mode (Fig.~\ref{f:schema}a, pink box) the clock is the
polarization of a photon. It is an extremely simple clock: it has a
dial with only two values, either $|H\>$ (detector 1 clicked)
corresponding to time $t = t_1$, or $|V\>$ (detector 2 clicked)
corresponding to time $t=t_2$.  [Here $t_2-t_1=\pi/2\omega$, where
$\omega$ is the polarization rotation rate of the quartz plate, since
the polarization is flipped in this time interval.] The experimenter
also measures the polarization of the first photon with detectors 3
and 4. This last measurement can be expressed as a function of time
(he has access to time only through the clock photon) by considering
the correlations between the results from the two photons: the
time-dependent probability that the first photon is vertically
polarized (i.e.~that detector 3 fires) is $p(t_1)=P_{3|1}$ and
$p(t_2)=P_{3|2}$, where $P_{3|x}$ is the conditional probability that
detector 3 fired, conditioned on detector $x$ firing (experimental
results are presented in Fig.~\ref{f:results}a). This type of
conditioning is typical of every time-dependent measurement:
experimenters always condition their results on the value they read on
the lab's clock (the second photon in this case). The experimenter has
access only to physical clocks, not to abstract coordinate time
\cite{w,ein,rovt}. In our experiment this restriction is implemented
by employing a different phase plate A (of random thickness unknown to
the experimenter) in every experimental run.

In super-observer mode (Fig.~\ref{f:schema}a, yellow box) the
experimenter takes the place of a hypothetical observer external to
the universe that has access to the abstract coordinate time and tests
whether the global state of the universe has any dependence on it.
Hence, he must perform a quantum interference experiment that tests
the coherence between the different histories (wavefunction branches)
corresponding to the different measurement outcomes of the internal
observers, represented by the which-way information after the
polarizing beam splitter PBS$_1$. In our setup, this interference is
implemented by the beam splitter BS of Fig.~\ref{f:schema}b. It is
basically a quantum erasure experiment \cite{scully,erasure} that
coherently ``erases'' the results of the time measurements of the
internal observer: conditioned on the photon exiting from the right
port of the beam splitter, the information on its input port (i.e.~the
outcome of the time measurement) is coherently erased \cite{preskill}.
The erasure of the time measurement by the internal observers is
necessary to avoid that the external observer (super-observer) himself
becomes correlated with the clock. However, the super-observer has
access to abstract coordinate time: he knows the thickness of the
blue plates, which is precluded to the internal observers, and he can
test whether the global state evolves (experimental results are
presented in Fig.~\ref{f:results}b).

In addition, we also test the GPPT mechanism showing that our
experiment can also account for two-time measurements (see
Fig.~\ref{f:schema}b). These are implemented by the two polarizing
beam splitter PBS$_1$ and PBS$_2$. PBS$_1$ represents the initial time
measurement that determines when the experiment starts: it is a
non-demolition measurement obtained by coupling the photon
polarization to its propagation direction, while the initialization of
the system state is here implemented through the entanglement. PBS$_2$
together with detectors 1 and 2 represents the final time measurement
by determining the final polarization of the photon. Between these two
time measurements both the system and the clock evolve freely (the
evolution is implemented by the birefringent plates A). In the GPPT
mechanism, the abstract coordinate time (the thickness of the quartz
plates A) is unaccessible and must be averaged over
\cite{gambinipullin,montev,pagereply}.  This restriction is
implemented in the experiment by avoiding to take into account the
thickness of the blue quartz plates A when extracting the conditional
probabilities from the coincidence rates: the rates obtained with
different plate thickness are all averaged together. The formal
mapping of the GPPT mechanism to our experiment is detailed in the
appendix.

\begin{figure}[tbp] \begin{center}
\includegraphics[width=0.4\textwidth]{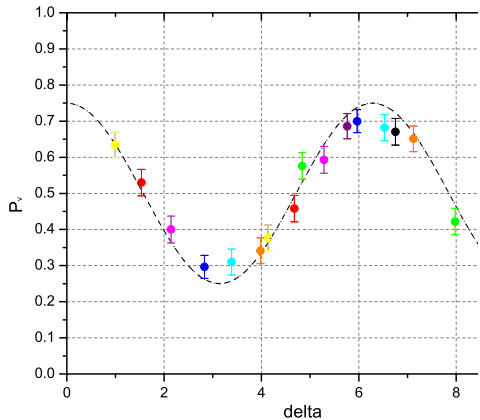}
\caption{GPPT experimental results: probability $p(t)$ that the
    upper photon is $V$ (namely that detector 3 clicked) as a function
    of the time $t$ recovered from the lower photon. The points with
    matching colors represent $p(t_1+\tau_i)$ and $p(t_2+\tau_i)$:
    yellow, red, blue, etc., for $i=0,1,2,\cdots$, respectively. Here
    nine different values of $\tau_i$ are obtained from a 1752$\mu$m
    thick quartz plate rotated by nine different angles from the
    vertical (14,16,18,20,21.5,23,25,27,29 degrees).  The dashed line
    is the theoretical value. Its reduced visibility is an expected
    effect of the use of imperfect clocks
    \cite{decoherence,montev,w}.}
\label{f:resultsgp} \end{center} \end{figure}

As before, the time dependent probability of finding the system photon
vertically polarized is $p(t_1)=P_{3|1}$ and $p(t_2)=P_{3|2}$.
However, a clock that returns only two possible values ($t_1$ and
$t_2$) is not very useful. To obtain a more interesting clock, the
experimenter performs the same conditional probability measurement
introducing varying time delays to the clock photon, implemented
through quartz plates of variable thickness (dashed box B in
Fig.~\ref{f:schema}b).  [Even though he has no access to abstract
coordinate time, he can have access to systems that implement known
time delays, that he can calibrate separately.]  Now, he obtains a
sequence of time-dependent values for the conditional probability:
$p(t_1+\tau_i)=P^{\tau_i}_{3|1}$ and $p(t_2+\tau_i)=P^{\tau_i}_{3|2}$,
where $\tau_i=\delta_i/\omega$ is the time delay of the clock photon
obtained by inserting the quartz plate B with thickness $\delta_i$ in
the clock photon path. The experimental results are presented in
Fig.~\ref{f:resultsgp}, where each colour represents a different delay:
the yellow points refer to $\tau_0$; the red points to $\tau_1$, etc.
They are in good agreement with the theory (dashed line) derived in
the appendix. The reduction in visibility of the
sinusoidal time dependence of the probability is caused by the
decoherence effect due to the use of a low-resolution clock (our clock
outputs only two possible values), a well known effect
\cite{decoherence,montev,w,cld}.

In summary, by running our experiment in two different modes
(``observer'' and ``super-observer'' mode) we have experimentally
shown how the same energy-entangled Hamiltonian eigenstate can be
perceived as evolving by the internal observers that test the
correlations between a clock subsystem and the rest (also when
considering two-time measurements), whereas it is static for the
super-observer that tests its global properties. Our experiment is a
practical implementation of the PaW and GPPT mechanisms but,
obviously, it cannot discriminate between these and other proposed
solutions for the problem of time
\cite{kuchar,isham,ashtekar,anderson,altra}. In closing, we note that
the time-dependent graphs of Fig.~\ref{f:resultsgp} have been obtained
without any reference to an external time (or phase) reference, but
only from measurements of correlations between the clock photon and
the rest: they are an implementation of a `relational' measurement of
a physical quantity (time) relative to an internal quantum reference
frame \cite{rudolphrmp,wiseman}.

\subsection*{Experimental setup}
The experimental setup (Fig.~\ref{f:exp}) consists of two blocks:
``preparation'' and ``measurement''.  The preparation block produces a
family of biphoton polarization entangled states of the form:
\begin{equation}
  \left|\Psi\right\rangle=\cos\theta\left|{HH}\right\rangle + 
e^{i\varphi} \sin\theta \left|{VV}\right\rangle
\label{eq:state}
\end{equation}
by exploiting the standard method of coherently superimposing the emission of two type I crystals whose optical axes are rotated of 90$^o$ \cite{m}.

The measurement block can be mounted in  different configurations corresponding to
``observer'' and ``super-observer'' ones of PaW and GPPT scheme (Fig.1). In general, each arm of
the measurement block contains interference filters (IF) with central
wavelength $702$ nm (FWHM $1$ nm) and a polarizing beam splitter (PBS).
Before the PBS the
polarization of both photons evolves in the birefringent quartz plates
A (blue boxes in Fig.~\ref{f:schema}) as $\left| V \right\rangle \to
\left| V \right\rangle\cos\delta +i\left| H \right\rangle\sin\delta$,
where $\delta$ is the material's optical thickness.
%Four avalanche photodiode detectors () are placed at the outputs
%and connected to a coincidence circuit (CC).  \\
\textbf{``Observer'' mode in PaW scheme} (Fig.~\ref{f:schema}, block a): 
In this mode, the polarization of the photon in the lower arm is used
as a clock: the first polarizing beam splitter PBS$_1$ acts as a
non-demolition measurement in the $H/V$ basis of the polarization of the
second photon, finally detected by  single-photon avalanche diodes  (SPAD) 
1, 2.  In this mode, the experimenter has no access to an
external clock, he can only use the correlations (coincidences)
between detectors: the time-dependent probability of finding the first
photon in $\left|V\right\rangle$ is obtained from the coincidence rate
between detectors 1-3 (corresponding to a measurement at time $t_1$),
or 2-3 (corresponding to a measurement at time $t_2$): appropriately
normalized, these coincidence rates yield the conditional
probabilities $P_{3|x}$. The impossibility to directly access abstract
coordinate time (the thickness of the plates) is implemented by
averaging the coincidence rates obtained for all possible thicknesses
of the birefringent plates A: the plate thickness does not enter into
the data processing in any way. \\
\textbf{``Super-observer'' mode in PaW scheme} (Fig.1b): This
mode is employed to prove that the global state is static with respect
to abstract coordinate time, represented by the thickness of the
quartz plates A. The $50/50$ beam splitter (BS) in block b performs a
quantum erasure of the polarization measurement (performed by the
polarizing beam splitter PBS$_1$) conditioned on the photon exiting
its right port. For temporal stability, the interferometer is placed
into a closed box.  The output state is reconstructed using
ququart state tomography \cite{kwiat2001,tomo,our} (the two-photon
polarization state lives in a four-dimensional Hilbert space), where
the projective measurements are realized with polarization filters
consisting of a sequence of quarter- and half-wave plates and a
polarization prism which transmits vertical polarization (Fig.4).  The fidelity between the tomographically
reconstructed state and the theoretical state $|\Psi\>$ is reported in
Fig.~\ref{f:results}b.
\textbf{GPPT two-time scheme} Here a second PBS preceding detectors allows
a two-time measurement. To obtain a more interesting time dependence than the
probability at only two times, we delay the clock photon with an
additional birefringent plate B (dashed box in Fig.~\ref{f:schema}), a 1752$\mu$m-thick
quartz plate rotated at nine different angles,
placed in the lower arm, and we repeat the same procedure described
above for different thicknesses of the plate B. This represents an
internal observer that introduces a (known) time delay to his clock
measurements. The results are shown in Fig.~\ref{f:results}.

\begin{figure}[tbp] \begin{center}
\includegraphics[width=0.4\textwidth]{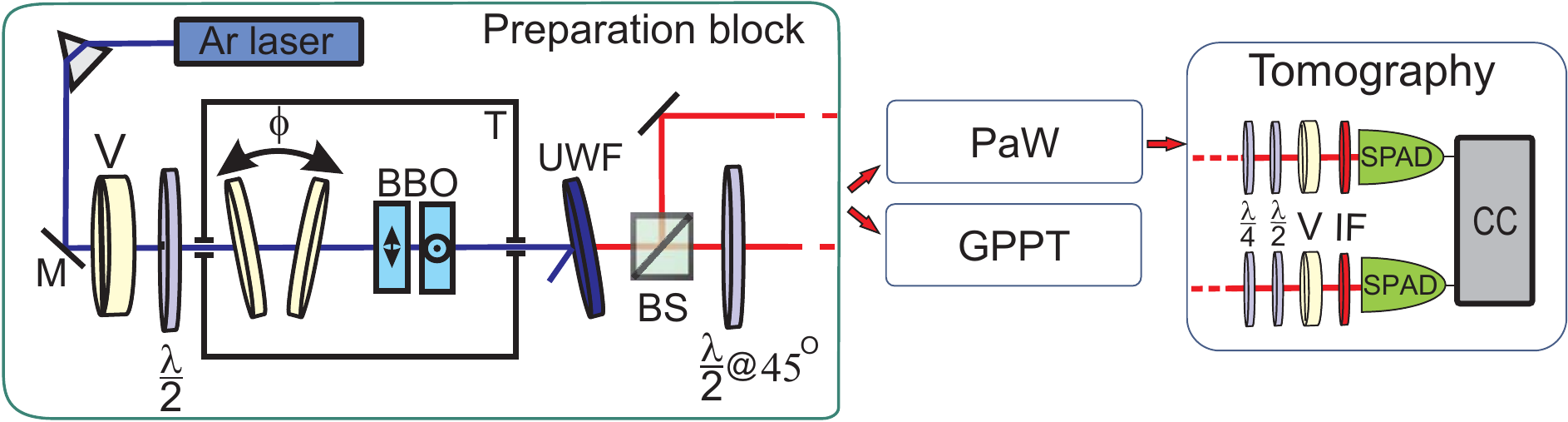}
\caption{ {\bf Preparation block}: Pairs of degenerate entangled
  photons are produced by pumping two orthogonally oriented type I BBO
  ($\beta-Ba B_2 O_4$) crystals (placed into a temperature-stabilized
  closed box T) pumped by a 700 mW Ar laser, later eliminated by a
  filter (UWF).  The basic state amplitudes are controlled by a
  Thompson prism (V), oriented vertically, and a half-wave plate
  $\lambda$/2 at angle $\theta$. Two 1 mm quartz plates, that can be
  rotated along the optical axis, introduce a phase shift ' between
  horizontally and vertically polarized photons.  The beam splitter
  (BS) is used to split the initial (collinear) biphoton field into
  distinct spatial modes. It prepares the singlet Bell state $\Psi$ of
  Eq. (1) (using $\theta = 45^o$; '$\phi= 0^o$, and an additional
  half-wave plate $\lambda$/2 at $45^o$ in the transmitted arm).  {\bf
    Measurement block}: We implement PaW or GPPT as in Fig.1. In PaW superobserver mode the final state is checked by quantum 
state tomography \cite{kwiat2001,tomo,our}, realised by registering the coincidence rate for
    16 different projections achieved through half and quarter wave plates and a
    fixed analyzer (V).  }
\label{F:exp} \end{center} \end{figure}

\subsection*{Appendix}
In this appendix we detail how our experiment
implements the Gambini et al. (GPPT) proposal
\cite{gambinipullin,montev} for extending the PaW mechanism
\cite{pw,w,pagereply} to describe multiple time measurements. We also
derive the theoretical curve of Fig.~\ref{f:resultsgp}.

Time-dependent measurements performed in the lab typically require two
time measurements: they establish the times at which the experiment
starts and ends, respectively. The PaW mechanism can accommodate the
description of these situations by supposing that the state of the
universe will contain records of the previous time measurements
\cite{pagereply}.  However, this observation in itself seems
insufficient to derive the two-time correlation functions (transition
probabilities and time propagators) with their required properties, a
strong criticism directed to the PaW mechanism
\cite{kuchar,pagereply}. The GPPT proposal manages to overcome this
criticism. It is composed of two main ingredients: the recourse to
Rovelli's `evolving constants' to describe observables that commute
with global constraints, and the averaging over the abstract
coordinate time to eliminate any dependence on it in the observables.
Our experiment tests  the latter aspect of the GPPT
theory.

Measurements of a physical quantity at a given clock time, say $t$,
are described by the conditional probability of obtaining an outcome
on the system, say $d$, given that clock time-measurement produces
the outcome $t$. This conditional probability is given by
\cite{gambinipullin,montev}
\begin{eqnarray}
  p(d|t)=
  \frac {\int dT\; \mbox{Tr}[P_{d,t}(T)
    \rho ]}{\int dT\;{\mbox{Tr}[P_{t}(T)\rho]}}
\labell{con}\;,
\end{eqnarray}
where $\rho$ is the global state, $P_{t}(T)$ is the projector relative
to a result $t$ for a clock measurement at coordinate time $T$ and
$P_{d,t}(T)$ is the projector relative to a result $d$ for a system
measurement and $t$ for a clock measurement at coordinate time $T$
(working in the Heisenberg picture with respect to coordinate time
$T$). Clearly, such expression can be readily generalized to arbitrary
POVM measurements. (A similar expression, but in the Schr\"odinger
picture, already appears in \cite{pagereply}.) The
integral that averages over the abstract coordinate time $T$ in
\eqref{con} embodies the inaccessibility of the time $T$ by the
experimenter: he can access only the clock time $t$, an outcome of
measurements on the clock system.

A generalization of this expression to multiple time measurements is
expressed by \cite{gambinipullin}
\begin{eqnarray}
\labell{gppt}  &&  p(d=d'|t_f,d_i,t_i)\\&&=
  \frac {\int dT\;\int dT'\; \mbox{Tr}[P_{d',t_f}(T)P_{d_i,t_i}(T')
    \;\rho\; P_{d_i,t_i}(T') ]}{\int dT\;\int
    dT'\;{\mbox{Tr}[P_{t_f}(T)P_{d_i,t_i}(T')\;\rho\; P_{d_i,t_i}(T')]}}
  \nonumber\;,
\end{eqnarray}
which gives the conditional probability of obtaining $d'$ on the
system given that the final clock measurement returns $t_f$ and given
that a ``previous'' joint measurement of the system and clock returns
$d_i$, $t_i$. (This expression can also be formulated as a
conventional state reduction driven by the first measurement
\cite{montev}.)

In our experiment to implement the GPPT mechanism
(Fig.~\ref{f:schema}b) we must calculate the conditional probability
that the system photon is $V$ (namely detector 3 clicks) given that
the clock photon is $H$ after the first polarizing beam splitter
PBS$_1$ (initial time measurement) and is $H$ or $V$ after the second
polarizing beam splitter (final time measurement). The initial time
measurement succeeds whenever one of photodetectors 1 or 2 click: this
means that the clock photon chose the $H$ path at PBS$_1$. (Our
experiment discards the events where the first time measurement at
PBS$_1$ finds $V$, although in principle one could easily take into
account these cases by adding a polarizing beam splitter and two
photodetectors also in the $V$ output mode of PBS$_1$.) The final time
measurement is given by the click either at photodetector 1 or 2: the
clock dial shows $t_f=t_1$ and $t_f=t_2=t_1+\pi/2\omega$, respectively.
Using the GPPT mechanism of Eq.~\eqref{gppt}, this means that the time
dependent probability that the system photon is vertical (detector 3
clicks) is given by
\begin{eqnarray}
\labell{sld}
  &&  p(d=3|t_f=t_k,d_i,t_i)\\&&=
  \frac {\int dT\;\int dT'\; \mbox{Tr}[P_{d=3,t_f=t_k}(T)P_{d_i,t_i}(T')
    \rho P_{d_i,t_i}(T') ]}{\int dT\;\int
    dT'\;{\mbox{Tr}[P_{t_f=t_k}(T)P_{d_i,t_i}(T')\rho P_{d_i,t_i}(T')]}}
  \nonumber\;,
\end{eqnarray}
where $P_{d=3,t_f=t_k}$ is the joint projector connected to detector 3
and detector $k=1$ or $k=2$ and $P_{d_i,t_i}$ is the projector
connected to the first time measurement. The latter projector is
implemented in our experiment by considering only those events where
either detector 1 or detector 2 clicks, this ensures that the clock
photon chose the $H$ path at PBS$_1$ (namely the initial time is
$t_i$) and that the system photon was initialized as $|V\>$ at time
$t_i$. (In principle, we could consider also a different initial time
$t'_i$ by employing also the events where the clock photons choose the
path $V$ at PBS$_1$.) Introducing the unitary abstract-time evolution
operators, $U_T$, the numerator of Eq.~\eqref{sld} becomes
\begin{eqnarray}
  \int dT\;\int dT'\; \mbox{Tr}[P_{d=3,t_f=t_k}U_{T-T'}
  P_{d_i,t_i}\:U_{T'}
  \rho U^\dag_{T'}P_{d_i,t_i}\times&&\nonumber\\U^\dag_{T-T'}]=\int dT\;
  \mbox{Tr}[P_{d=3,t_f=t_k}U_{T}
  P_{d_i,t_i}
  \rho P_{d_i,t_i}U^\dag_{T}],&&
\nonumber\;
\end{eqnarray}
where we use the property $U_T {U_{T'}}^\dag=U_{T-T'}$ and we dropped
one of the two time integrals by taking advantage of the time
invariance of the global state $\rho$ (which has been also tested
experimentally in the super-observer mode). Gambini et al. typically
suppose that the clock and the rest are in a factorized state
\cite{montev}, but this hypothesis is not strictly necessary for their
theory \cite{gambinipullin}: we drop it so that we can use the same
initial global state that we used for testing the PaW mechanism.

Using the same procedure also to calculate the denominator of
Eq.~\eqref{sld}, we can rewrite this equation as
\begin{eqnarray}
\labell{sld1}
  p(d=3|t_f=t_k,d_i,t_i)=
  \frac {\mbox{Tr}[P_{d=3,t_f=t_k}\bar\rho
    ]}{\mbox{Tr}[P_{t_f=t_k}\bar\rho]}
\;,
\end{eqnarray}
where $\bar\rho$ is the time-average of the global state after the
first projection, namely
\begin{eqnarray}
  \bar\rho\propto\int dT\:U_{T}
  \rho_{t_i,d_i}U^\dag_{T}\;,\qquad   \rho_{t_i,d_i}\equiv P_{d_i,t_i}
  \rho P_{d_i,t_i}
\labell{rhoba}\;,
\end{eqnarray}
where the averaging over the abstract coordinate time $T$ is used to
remove its dependence from the state. In our experiment such average
is implemented by introducing random values of the phase plates A
(unknown to the experimenter) in different experimental
runs.

In our GPPT experiment there are two possible values for the initial
projector $P_{d_i,t_i}$: either the clock photon is projected on the
$H$ path after PBS$_1$ (corresponding to an initial time $t_i$) or it
is projected onto the $V$ path (corresponding to an initial time
$t_i+\pi/2\omega$). We will consider only the first case, which
corresponds to a click of either detector 1 or 2: we are
post-selecting only on the experiments where the initial time is
$t_i$. In this case, the global initial state will be $|H\>_c|V\>_r$
which is evolved into  the vector
$|\Psi(T)\>=[\cos(\omega(T+\tau))|H\>_c-
\sin(\omega(T+\tau))|V\>_c]
[\cos \omega T|V\>_r+\sin \omega T|H\>_r]$
where $\cal H$ is the global Hamiltonian defined in the main text and
$\tau$ is the time delay introduced by the plate B of
Fig.~\ref{f:schema}b. Moreover, the projectors in Eq.~\eqref{sld1} are
\begin{eqnarray}
&&\nonumber
P_{d=3,t_f=t_k}\equiv|k\>_c\<k|\otimes|V\>_r\<V|\;,\mbox{ and }\\&&
P_{t_f=t_k}\equiv|k\>_c\<k|\otimes\openone_r
\labell{proj}\;,
\end{eqnarray}
where $|k=0\>_c\equiv|H\>_c$ and $|k=1\>_c\equiv|V\>_c$. The projector
$P_{d=3,t_f=t_k}$ corresponds to the joint click of detectors $k$ and
3, while $P_{t_f=t_k}$ corresponds to the click of detector $k$ and
either one of detectors 3 or 4. In other words, Eq.~\eqref{sld1} can
be written as
\begin{eqnarray}
  p(d=3|t_f=t_k,d_i,t_i)=P_{3k}/(P_{3k}+P_{4k})
\labell{sld2}\;,
\end{eqnarray}
where $P_{jk}$ is the joint probability of detectors $j$ and $k$
clicking. For example, $P_{32}$ is the joint probability that detector
3 and 2 click, namely that both the clock and the system photon were
$V$. Considering only the component $|V\>_c|V\>_r$ of the state
$|\Psi(T)\>$,  this is given by
\begin{eqnarray}
P_{32}=\tfrac1{2\pi}\int_0^{2\pi}
\!\!d\varphi\sin^2(\varphi+\omega\tau)\cos^2\varphi=\frac{1+2\cos^2\omega\tau}8
\labell{p32}\;
\end{eqnarray}
where we have calculated the integral over $T$ of Eq.~\eqref{rhoba}
using a change of variables $\omega T=\varphi$.  Proceeding
analogously for all the other joint probabilities, namely replacing
the projectors \eqref{proj} into \eqref{sld1}, we find the probability
for detector 3 clicking (namely the system photon being $V$)
conditioned on the time $t_f$ read on the clock photon as
\begin{eqnarray}
p(3|t_f=t_1)&=&(1+2\cos^2\omega\tau)/4\labell{l2}\\
p(3|t_f=t_2)&=&(1+2\sin^2\omega\tau)/4
\labell{l1}\;,
\end{eqnarray}
which is plotted as a function of $\tau$ in Fig.~\ref{f:results}b
(dashed line). Since $t_2=t_1+\pi/2\omega$, we have plotted the points
relative to $p(3|t_2)$ as displaced by $\pi/2$ with respect to the
points relative to $p(3|t_1)$, so that the two curves \eqref{l2} and
\eqref{l1} are superimposed in Fig.~\ref{f:results}.

%In closing, we also point out an alternative, perhaps more intuitive
%way to argue for the staticity of the state of the universe in
%addition to the Wheeler-De Witt equation. In a sense, this staticity
%is a trivial application of Mach's principle (in cosmologies where
%that is possible): a synchronous time shift of the whole universe must
%be irrelevant since there is no external time reference relative to
%which such shift could be detected, hence the universe as a whole must
%be static \cite{pw}. Page and Wootters' proposal naturally
%embodies the philosophy of relationalism \cite{rovelli,rovt} and
%operationalism, since time is only defined in relation to clocks and
%to its measurement procedure \cite{rudolphrmp,wiseman}.

\section*{Acknowledgments}
We thank A. Ashtekar for making us aware of Ref.~\cite{gambinipullin}.
We acknowledge the Compagnia di San Paolo for partial support.E.V.Moreva acknowledges the support from the Dynasty Foundation and Russian Foundation for Basic Research (project 13-02-01170-D)

\end{document}